\documentclass[preprint]{aastex}

\usepackage{ifpdf}
\usepackage{graphicx}
\usepackage{natbib}

\usepackage{amsmath}
\usepackage{amssymb}
\usepackage{float}
\usepackage{graphics}
\usepackage{graphicx}
\newcommand{\mpl}{m_{\mathrm{pl}}}
\newcommand{\dd}{\, {\rm d}}
\newcommand{\gsim}{\;\mbox{\raisebox{-0.5ex}{$\stackrel{>}{\scriptstyle{\sim}}$}
}\;}
\newcommand{\lsim}{\;\mbox{\raisebox{-0.5ex}{$\stackrel{<}{\scriptstyle{\sim}}$}
}\;}

\begin{document}

\title{Astrophysical Tests of Modified Gravity: Constraints from Distance Indicators in the Nearby Universe}

\author{Bhuvnesh Jain\altaffilmark{1}, Vinu Vikram\altaffilmark{1}, Jeremy
Sakstein\altaffilmark{2}}
\bibliographystyle{apj}

\altaffiltext{1}{Department of Physics \& Astronomy, University of
Pennsylvania, Philadelphia, PA 19104, USA}
\altaffiltext{2}{Department of Applied Mathematics and Theoretical
  Physics, Cambridge, CB3 0WA, UK}

\begin{abstract}
We use distance measurements in the nearby universe to carry out new tests of  
gravity, surpassing other astrophysical tests by over two orders of magnitude for chameleon theories. The three  nearby distance indicators -- cepheids, tip of the red giant branch (TRGB) stars, and water masers -- operate in gravitational fields of widely different strengths. 
This enables tests of scalar-tensor gravity theories because they are screened 
from enhanced forces to different extents. 
Inferred distances from cepheids and TRGB stars are altered (in opposite directions) over a range of chameleon gravity theory parameters well below the sensitivity of 
 cosmological probes. Using published data we have compared cepheid and TRGB distances in a sample of unscreened dwarf galaxies within 10 Mpc. As a control sample we use a comparable set of screened galaxies. We find no evidence for the order unity force enhancements expected in these theories. Using a two-parameter description of the models (the coupling strength and background field value) we obtain constraints on both the chameleon and symmetron screening scenarios. In 
particular we show that $f(R)$ models with background field values 
$f_{R0}$ above $5\times 10^{-7}$ are ruled out at 
the 95\% confidence level. We also compare TRGB and maser distances to the galaxy NGC 4258 as a second test for larger field values. 
While there are several approximations and caveats in our study, our analysis 
demonstrates the power of gravity tests in the local universe. We discuss the prospects
for additional, improved tests with future observations. 
\end{abstract}

\maketitle


\section{Introduction}
\subsection{Modified Gravity}

Modified theories of gravity (MG) have received a lot of attention in
recent years. Several unexplained phenomena such as the 
observed accelerated expansion of the universe, spatio-temporal variation of
the fundamental constants (e.g. the fine-structure constant) and dark
matter can in principle be explained by modifying general relativity (GR) on
large (astrophysical and greater) scales. This study is motivated by recent work 
on MG to obtain cosmic acceleration without invoking quintessence-like dark energy. 
Even on theoretical grounds, GR is unlikely to be
the complete theory of gravity owing to 
singularities and its non-renormalizibility. Hence it is generally
considered to be the low energy effective action of some UV-complete
theory (though our interest is modifications in the long distance/low energy regime). 

Any modification to GR generically introduces an additional
degree of freedom (Weinberg 1965) and scalar-tensor theories, where a
new scalar field couples non-minimally to gravity, are ubiquitous in
many attempts to find consistent theories. %
Theories of higher dimensional gravity often appear as scalar-tensor theories to observers in four dimensions. Theories
such as DGP (Dvali, Gabadadze \& Porrati 2000) invoke 
braneworld scenarios  -- the
distance between two branes acts as the scalar field. 
In Kaluza-Klein models, the parameters controlling the size of the compact directions appear as a scalar coupled to gravity in four dimensions. 
The low energy
effective action of string theory contains a new scalar particle, the
dilaton, coupled non-minimally to gravity and any theory with
$\mathcal{N}=4$ supergravity (or higher) contains at least two scalar
fields in the gravity multiplet. Even attempts to modify the
geometrical properties of GR tend to lead to scalar-tensor theories;
for example the entire class of $f(R)$ theories is equivalent to a
single scalar field coupled non-minimally to matter via a Weyl
rescaling of the metric. 

 Thus  a wide class of gravity theories contain a coupling of the scalar field to
matter via a universal fifth force which leads to enhancements of the gravitational force.
Non-relativistic matter -- such as the stars, gas, and dust in galaxies --
will feel this enhanced force and as a consequence,
a general feature of scalar-tensor theories is that dynamically
inferred masses are larger than the true masses.
The discrepancy can be up to a factor of $1/3$ in $f(R)$
or DGP gravity 
(for recent reviews see Silvestri \& Trodden 2009 and  Jain \& Khoury 2010). 
Photons do not feel the enhanced force, so that lensing
probes the true mass distribution.


This enhanced gravitational force should be detectable through
fifth force searches such as the E\"{o}t-Wash experiment (Kapner et al. 2007) and Casimir force experiments (e.g. Decca et al 2003) as well as tests of the equivalence principle\footnote{If the scalar coupling to matter is not constant then one generically expects violations of the weak equivalence principle.} (Mota \& Shaw 2006) or through the orbits of planets around the Sun (Will 2006).
However, since all of these experiments have been performed in our
local vicinity, i.e. the solar system, they do not rule out any
large-scale modifications where fifth forces are active over large
(cosmological) scales while matching  the predictions of GR within experimental 
bounds on small scales. 
Any theory where the fifth forces are suppressed on
small scales is said to possess a \textit{screening mechanism}:
regions where fifth forces are active are said to be
\textit{unscreened} whereas those where they are suppressed are
\textit{screened}. Khoury \& Weltman (2004) proposed such a
mechanism where non-linear screening of the scalar field, known as
\textit{chameleon screening}, can suppress the fifth force
in high density environments such as the Milky Way, so that solar system
and laboratory tests can be satisfied.

In this paper we will consider deviations from GR exhibited in
theories that rely on chameleon screening.
Qualitatively similar behavior occurs in symmetron screening
(Hinterbichler \& Khoury 2010) and the environmentally dependent
dilaton (Brax et al. 2010) and the tests we will present here apply
 to these mechanisms as well
 We also note that chameleon screening was originally suggested
to hide the effects of a quintessence-like scalar that forms the dark energy
and may couple to matter (generically such a coupling is expected unless
forbidden by a symmetry). Hence there are reasons to expect such a screening
effect to operate in either a dark energy or modified gravity scenario, or even
in scenarios that don't relate to cosmic acceleration such as the scalar
fields invoked in string theories.

A different screening mechanism, Vainshtein screening (Vainshtein 1972), operates
by including non-canonical kinetic terms in the field equations whose
non-linear nature acts to recover GR on scales smaller than some
\textit{Vainshtein radius}. In theories that contain this mechanism
(such as DGP, massive gravity and Galileons) this radius must
typically be taken to be of the order of the length scale of typical
galaxies, independent of their mass, and so does not produce the
observable effects considered here (see Appleby and Linder 2012 for recent
cosmological constraints for a subclass of Galileon theories). 
However recent studies
have pointed out the possibility of enhanced forces even within galaxies 
in Vainshtein theories, opening
them up to  astrophysical tests  (Chan \& Scoccimarro 2009; Hui \& Nicolis 2012).

\subsection{Observations of the Nearby Universe}

The logic of screening of the fifth force in scalar-tensor
gravity theories implies that signatures of modified gravity will
exist where gravity is weak. In particular,
dwarf galaxies in low-density environments may remain unscreened as the
Newtonian potential $\Phi_{\rm N}$, which determines the level of screening,
is at least an order of magnitude smaller than in the Milky Way. Hence
dwarf galaxies can exhibit manifestations of modified forces in both
their infall motions and internal dynamics.
Hui, Nicolis \& Stubbs (2009)
and Jain \& Vanderplas (2011) discuss various observational effects
while Vikram et al (2012, in preparation) present a set of tests from current observations.

Stars within unscreened galaxies may show the effects of modified gravity.
Chang \& Hui (2010) and Davis et al. (2011a) describe the effects on  giant
and main sequence stars, respectively: essentially the enhanced gravitational force
makes stars of a given mass brighter and hotter than in GR. They are also more ephemeral since they consume their fuel at a faster rate.
For the Sun the  potential  $\Phi_{\rm N} \approx 2\times 10^{-6}$ (we set $c=1$ and work with the amplitude of the potential throughout).
Coincidentally, the potential of the Milky Way is close to this value -- and is
believed to be sufficient to screen the galaxy so that solar system tests of
gravity are satisfied
\footnote{At least this is the straightforward interpretation; there may be
loopholes to this logic where the galaxy is screened by the Newtonian
potential of the local group or structure on larger scales}.
Thus main sequence stars whose masses are similar to that of the Sun
are likely to be partially or completely screened. It is worth noting
that these screening conditions are found by considering static,
non-dynamical spherical objects sitting inside a fixed scalar
field. Davis et al. (2011a) have shown that stars, as
dynamical objects which support themselves under the action of the
enhanced gravity, can be partially unscreened at Newtonian
potentials where this simple model would predict them to be completely
screened, especially for higher mass objects.

Red giants are at least ten times larger in size than the main
sequence star from which
they originated so they have $\Phi_{\rm N} \sim 10^{-7}$ -- thus their envelopes
may be unscreened. The enhanced forces lead to smaller radii, hotter
surface temperatures and higher luminosities than their GR doppelgangers.
For $f(R)$ theories  (Capozziello et al 2003; Nojiri \& Odintsov 2003; Carroll et al 2004; 
Starobinsky 2007; Hu \& Sawicki 2007)
with background field
values ($f_{R0}$) in the
range $10^{-6}$--$10^{-7}$,
Chang \& Hui (henceforth CH) find that compared to GR a solar mass red
giant has radius $R$ smaller by over 5\%, luminosity $L$ larger by over
50\% at fixed effective temperature $T_e$, while
$T_e$ itself is higher by about 5\% at fixed $L$. They point
out that the change in surface temperature of about 150 Kelvin may be detectable
from data on the red giant branch in observed Hertzsprung-Russel (HR) diagrams.

We focus on specific stages of the evolution of  giants and supergiants to
seek different observational signatures. The first feature, well known
for its use in obtaining distances, is the nearly universal
luminosity of  $\lsim 2 M_\odot$ stars at the tip of the red giant branch. 
The second is the
Period-Luminosity
relation of cepheids, which are giant stars with $\sim 3-10 M_\odot$ that
pulsate when their
evolutionary tracks cross a near universal, narrow range in $T_e$ known as the
\textit{instability strip}.
The tight relation between luminosity and other observables is what makes these stars valuable distance indicators -- it also
makes them useful for tests of gravity.
For background field values  in the range $10^{-6}$--$10^{-7}$, the
 TRGB luminosity is largely robust to modified gravity while the cepheid $P-L$
relation is altered. Measurements of these properties
within screened and unscreened galaxies then provide tests of gravity:
the two distance indicators should agree for screened galaxies but not for
unscreened galaxies.

This paper is organised as follows: In \S 2 we briefly describe chameleon gravity (a full discussion for the interested reader is given in Appendix A) and explain the differences in stellar evolution due to its influence. In \S 3 the properties
of cepheids in GR and MG are presented, as well as a summary of relevant observations; the corresponding details of TRGB distances are presented in \S 4.
\S 5 contains our main results based on a comparison of cepheid and TRGB distances. 
Water masers in NGC 4258 are used for a second tests in \S 6, and other distance indicators are discussed.  We conclude in \S 7.

\section{Modified Gravity and its Effect on Stellar Structure}

\subsection{Review of Chameleon Screening}

Here we will briefly review the parameters that provide tests of
gravity, motivated by the chameleon-like screening mechanisms.
We refer the interested reader to Appendix A for the full details and
examples of some of the more common models.

There are two parameters in these theories. The first, $\chi_c$ (see Appendix A
and Davis et al. 2011a) determines how efficient the body is at screening
itself. If the magnitude of the surface Newtonian potential $\Phi_{\rm N} \ll \chi_c$ then
the object will be completely unscreened whilst if the converse is
true then the body will be at least partially screened (see equation \ref{eq:screenrad} in appendix A). 
Currently
there are two different constraints on $\chi_c$ in the literature. The
Newtonian potential of the Sun is around $2\times 10^{-6}$ and the
Milky way has a similar value and so if one demands that these objects
self screen then $\chi_c\gsim 10^{-6}$ is ruled out
observationally. 
Independent constraints come from cosmological observables and 
cluster abundances and set
an upper bound $\chi_c\approx 10^{-4}$ for $f(R)$ models 
(Schmidt, Vikhlinin \& Hu 2009;
Lombriser et al 2010 and references therein). 
Our analysis here constitute an independent
constraint, so we examine several possible uses of TRGB,
cepheid and water maser distance indicators to test gravity. 

The second parameter in these models is $\alpha_c$, which sets the
strength of the fifth force in unscreened regions. A completely
unscreened object will feel a fifth force that is proportional to the
Newtonian force and the combined forces simply amount to a rescaling
of $G$ by
\begin{equation}
G\rightarrow G(1+\alpha_c) .
\end{equation}
For partially screened
objects, the total force in the region exterior to the screening
radius can be described by a position dependent rescaling of G:
\begin{equation}\label{eq:g(r)}
G(r)=G\left[1+\alpha_c\left(1-\frac{M(r_{\rm s})}{M(r)}\right)\right]
\end{equation}
where $M(r)$ is the mass interior to a shell of radius $r$.

We will consider tests of chameleon theories that probe ranges of 
$\chi_c$ and $\alpha_c$ well below current astrophysical limits. 
For concreteness in evaluating screening conditions 
we work with $f(R)$ models, which are chameleon models 
with parameters:
\begin{equation}
\alpha_c=1/3; \ \ \chi_c=f_{R0} .
\end{equation}
where
$f_{R0}$ is a parameter commonly used in the literature to constrain
the model (Hu \& Sawicki 2007). $\alpha_c=1/3$ is also the value found
in the high density limit of the environmentally dependent dilaton
(see Appendix A). We will also consider other values of $\alpha_c$
in comparisons with observations.


In $f(R)$ theories, the parameter $f_{R0}$ sets the screening condition for an isolated spherical halo (Hu \& Sawicki 2007):
$\Phi_{\rm N} > \frac{3}{2} f_{R0}$ .
The Navarro-Frenk-White (NFW) density profile is a good approximation
for halos of most galaxies both in GR and $f(R)$ theories (Lombriser et al. 2012). It is useful to express the 
Newtonian potential in terms of the scale radius $r_s$ of the NFW profile as:
\begin{equation}
\Phi_{\rm N}^{NFW} = \frac{G M_g}{r}\ {\rm ln}(1+r/r_s)
\end{equation}
where we have followed Hu \& Sawicki (2007) in using $M_g$ to represent  the mass contained within $5.3 r_s$ (see also Schmidt 2010).
 The screening condition can then be conveniently expressed in terms of the observable maximum  circular velocity $v_{\rm max}$ as:
 \begin{equation}
 \left(\frac{v_{\rm max}}{100\ km/s}\right)^2 \gsim \frac{f_{R0}}{2\times10^{-7} } \ .
 \label{eqn:screening_vmax}
 \end{equation}
For the late-type dwarf galaxies of interest here, the peak circular velocity
is likely to be reasonably well estimated by the observed rotation curves. The
effects of inclination and of limited radial coverage typically lead to
$10\%$ level underestimates of $v_{\rm max}$. For the range of circular
velocities used in our sample, we probe
$f_{R0}$ values in the range $10^{-6}-10^{-7}$. 

\subsection{Stellar Structure and Evolution}

The structure of a spherically symmetric star is obtained by solving the equations
of stellar structure that at a given radius $r$ relate $M(r)$ to
 $P(r)$, $\rho(r)$ and $T(r)$ -- respectively the pressure,
density and temperature. The opacity $\kappa$  and the energy generation
rate $\epsilon$ are needed to close the equations. We will denote  the radius of the star as $R$,
so that  $M=M(R)$
is the total mass and $L=L(R)$ the total luminosity. $P$, $\kappa$ and
$\epsilon$ are determined in terms of $\rho$ and $T$ by equations of
state which are independent of gravitational physics. As noted by CH and Davis
et al. 2011a, modifying gravity only alters the gravitational physics, which is entirely contained in the equation of hydrostatic equilibrium:
\begin{equation}
\frac{\dd P}{\dd r} = -\frac{G(r) \rho(r) M(r)}{r^2},
\end{equation}
which represents the condition for the outward pressure to balance the (now enhanced) inward gravitational pull and remain static.
Note that the modification is expressed purely as a change in $G$,
which becomes dependent on $r$ if the star is partially screened
according to equation \ref{eq:g(r)}. The other three equations -- the
continuity, radiative transfer and energy generation equations -- are all
unaffected by this change in $G$. The result is that unscreened
stars of a given mass are more compact, brighter, and have a larger
effective temperature than screened stars of identical mass and
chemical composition. They also have a shorter main
sequence life-time due to their increased burning rate and finite fuel
supply.

We will assume that changes in the gravity theories occur on timescales longer than
the evolutionary timescales of stars. Since we are interested in massive post-main sequence stars,
evolutionary timescales are shorter than a billion years, typically orders of magnitude smaller.
Thus provided the main sequence is the same as in GR (i.e. for $f_{R0} < 10^{-6}$),
the star's post-main sequence evolution begins at the same point in
the HR diagram as in GR, and subsequently responds to a static (but
possibly $r$-dependent) $G$. Our results do not require this assumption but 
it simplifies the story. 

The complete system of stellar structure equations for main sequence
stars can be solved under certain simplifying assumptions (see Davis
et al. 2011a). However, if one really wants to look at the dynamical
and nuclear properties, as well as the structure of post main sequence
stars then a numerical prescription is needed. In this work we shall
use a modification of the publicly available stellar evolution code
MESA (Paxton et al. 2011) presented in (Davis et al. 2011a). MESA can simulate
individual stars of any given initial mass and metallicity and
includes a fully consistent implementation of nuclear reaction
networks, mass loss, convection, radiative transfer, opacity tables
and metallicity effects. In brief, it is a one-dimensional (in that it
assumes spherical symmetry) code that divides the star into radial cells of
unequal width, each with a set of properties such as mass, temperature
etc. The implementation of chameleon gravity into MESA uses a
quasi-static approximation where the structure of the star is solved
first and then the value of $G$ is updated in every cell given this
structure. This approximation is good provided that the time between
successive stellar models is smaller than the timescale over which the
changes in $G(r)$ are significant and MESA provides a facility to
ensure that this is always the case. This method has been checked
against that used by CH, who use a scalar-field ansatz and
interpolation between different cells, and the results have been found
to be indistinguishable. Our method uses the general screening
properties set out in Appendix A and as such, applies to generic  
screening behavior, not only chameleons.

The implicit relation for the screening radius is given in appendix A (equation \ref{eq:screenrad}) and is completely equivalent to the
condition
\begin{equation}\label{eq:starscrrad}
\chi_c=4\pi G\int_{r_{\rm s}}^R r\rho(r)\dd r.
\end{equation}
Given an initial stellar model, the integral \ref{eq:starscrrad} is performed from the stellar surface, cell by cell until the condition is satisfied. The cell where this is so is then designated as the screening radius and the value of $G$ is changed according to equation \ref{eq:g(r)} in all cells exterior to this. The star is then allowed to evolve under this new gravity until the next stellar model is reached and the process is repeated.

\section{Cepheid Variables}

\begin{figure}[t,h]
\centering
\includegraphics[width=10cm]{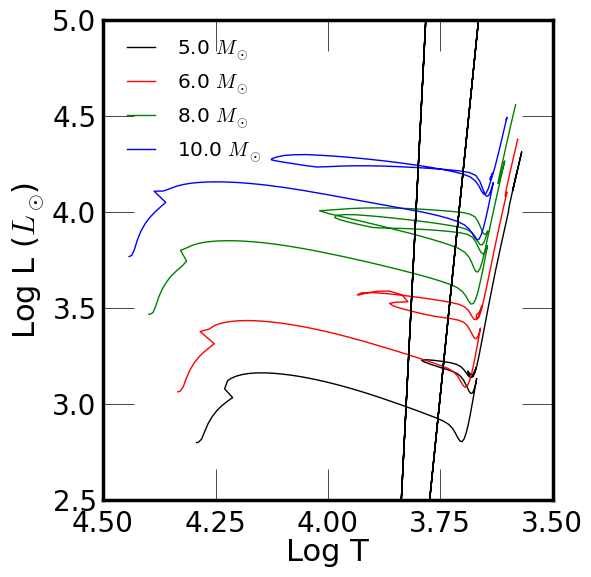}
\caption{The post-main sequence HR diagram ($L$ vs. $T_{e}$)
is shown for stars with different masses as indicated.
The black vertical lines shows the blue
edge (left) and red edge (right) of the instability strip. At each crossing of the instability strip the star pulsates as a cepheid variable. The fits for the instability strip
are taken from Alibert et al 1999. }  
\label{fig:T-L-mass}
\end{figure}

\subsection{Cepheid Pulsations in GR}

Cepheids are a class of massive, pulsating giant stars whose time period $P$ 
is related to luminosity $L$ in a well understood way. 
After a  giant reaches the tip of the red giant branch, its luminosity falls
and it
moves into the core Helium burning stage. Stars that are sufficiently massive, $M\gsim 
3 M_\odot$,
actually follow trajectories called loops, which at three or more phases cross
the instability strip -- set by a narrow range in effective temperature $T_e$. 
While in the instability strip, their luminosity and radius oscillate with a time
period of days-weeks in a very regular manner. 
We are concerned here with a class of pulsating stars called
(classical)
cepheids. The radial oscillations in a cepheid are the result of acoustic waves
resonating within the star. For much of the lifetime of a star, its envelope is stable
to pulsations -- the exception is the instability strip. Figure \ref{fig:T-L-mass} shows 
the post main sequence evolution of stars of different masses. 

The instability strip arises due to the presence of a He$^+$ ionisation zone in the stellar envelope located where the temperature is around $4\times10^4$ K (corresponding to the ionisation potential of He). The pulsation is driven by the $\kappa$-mechanism (and to a small extent the $\gamma$-mechanism\footnote{This is where the energy absorbed from the radiation is used to ionise the He and not to raise the temperature and so small compressions raise the opacity by virtue of the increased density, the opposite of their usual effect.}). The opacity throughout the majority of the star  is given by Kramer's law, $\kappa(R)\propto\rho T^{-3.5}$. In the ionisation zone however, the gradient $\dd \ln \kappa/\dd \ln T\gg -3.5$ and so a small compression, which increases the temperature slightly, causes a large increase in the opacity, absorbing radiation and damming up energy. 
This further increases the opacity, resulting in an outward pressure which drives the pulsations. This driving is only really effective when the thermal time-scale of the zone is comparable to the pulsation period, which requires it to be located in the so called \textit{transition region}, where the stellar processes (which are adiabatic in the star's interior) are becoming non-adiabatic. 
The instability strip therefore corresponds to the region in the H-R diagram where the ionisation zone coincides with this transition region. 

Stars can cross the instability strip multiple times. The 1st crossing of the
instability strip is before the star goes up the red giant branch and is far too brief to be  observationally
irrelevant. The 2nd crossing of the instability strip is the first crossing
after the tip of the red giant branch when the star is on the lower part of the
blue loop. And the 3rd crossing is when it is on the upper part of the blue
loop. The 2nd and 3rd crossing of the strip, and in particular on the blue edge,
is probably where cepheid properties are best understood. 
There are nearly as many observed
cepheids on the 2nd and 3rd crossing. We will use the 3rd crossing of a 6 
solar mass star as our fiducial case (observed cepheids are typically 6-8 solar 
mass stars). 
 
To estimate the pulsation period, one needs to go beyond hydrostatic equilibrium and consider
the full dynamical radial acceleration of a fluid element at radius $r$, which is described by the momentum equation:
\begin{equation}
\ddot{r} = -\frac{GM(r)}{r} - \frac{1}{\rho}\frac{\partial{P}}{\partial{r}}
\label{eqn: acceleration}
\end{equation}
The time period of pulsations $\Pi$ may
be estimated through various approximations:
1. As the sound crossing time for the diameter of the star, giving
 $\Pi \propto 1/\sqrt{\gamma G \rho}$ where $\gamma$ is the ratio of specific
heats and $\rho$ is the mean density of the star. This expression assumes that
physical variables like the density and
 temperature can be used. The dependence on $G$ and $\rho$ is essentially
correct, but the dependence on $\gamma$ is not correct.
 2. By perturbing equation \ref{eqn: acceleration} as well as the equations of continuity, radiative transfer and energy generation one can derive a full non-adiabatic wave equation for infinitesimal fluid elements in the Lagrangian picture (see Cox 1980). If one linearises this equation in the adiabatic limit and searches for standing wave solutions then the resultant eigenvalue equation for the radial wave, the linear adiabatic wave equation (LAWE), gives the pulsation periods. 
 The LAWE is highly non-trivial and depends on the zeroth order pressure and first adiabatic index $\Gamma_1$ and so the general case requires numerical matrix or shooting methods applied to simulations involving envelope models. One simplifying assumption that can be made however is that of a static sphere of fixed equilibrium radius and constant density composed of gas with an adiabatic relation 
 $\delta P/P \propto \gamma\ \delta \rho/ \rho$. Under this assumption, the LAWE can be solved for the period of small oscillations (note that the linear and adiabatic nature precludes a calculation of the amplitude) to find:  
\begin{equation}
\Pi = \frac{2\pi}{\sqrt{4/3 \pi G \rho (3\gamma-4)}} .
\label{eqn:period1}
\end{equation}
Using $\gamma=5/3$ for an ideal monatomic gas yields $\Pi = \sqrt{3\pi /
G\rho}$
$\Pi$ above is in the range of 1-100 days for a range of relevant red giant
densities.
This equation gives values of $\Pi$ that are fairly close to more detailed
calculations
as well as observations. 

We note that detailed numerical models are able to predict not just the period
but many detailed properties of cepheids, including the variations of size and 
luminosity, the location of the instability strip, and 
the dependence on mass and metallicity (e.g. Bono et al 1998). 
The primary source of uncertainty is the treatment of convection, so to some 
extent input from observations is used in theoretical models. 


\begin{figure}[t,h]
\centering
\includegraphics[width=8cm]{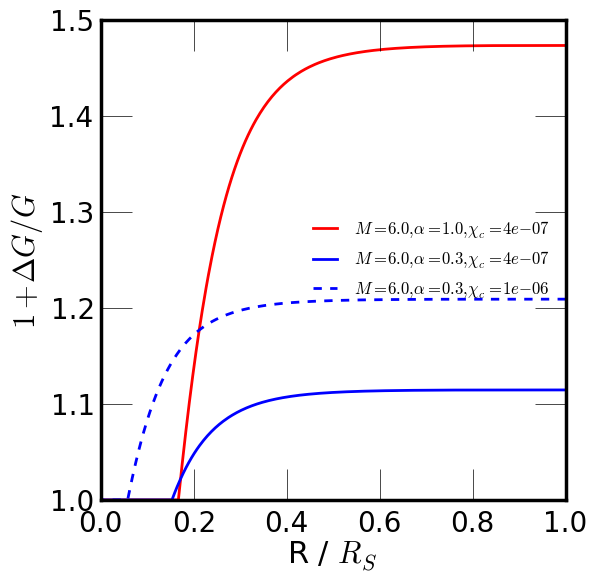}
\includegraphics[width=8cm]{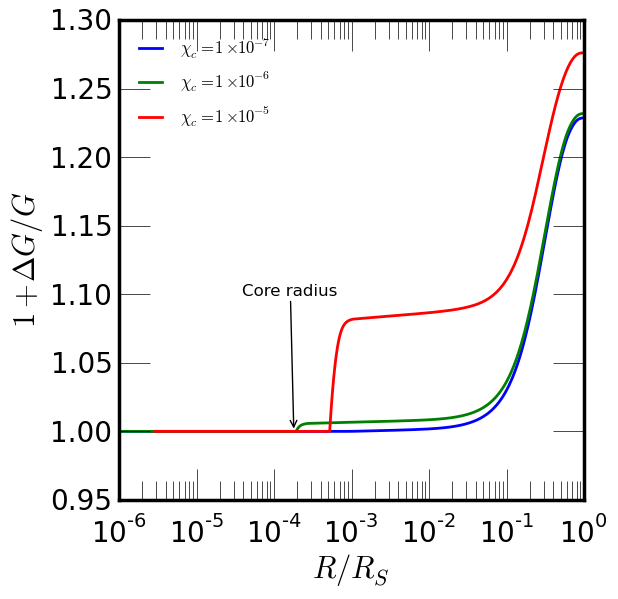}
\caption{
The profile of the effective gravitational constant $G$ inside the star
for cepheids (left panel) and TRGB stars (right panel). 
Note that
cepheid pulsations  
typically span $0.3 < R/R_S < 0.9$ while the core physics that
sets the TRGB luminosity occurs deep within the star. The core radius
in the right panel is shown by the arrow for $\chi_c=10^{-5}$. 
The plateau just below 1.1 on the y-axis shows the enhanced
gravity in the H-burning shell. } 
\label{fig:G-profiles}
\end{figure}

\subsection{Cepheid Pulsations in Modified Gravity}

The effect of MG theories on cepheid pulsations  is well approximated by 
considering the deviation of  
$G$ from its Newtonian value $G_N$.  
We apply Eqn. \ref{eqn:period1} to estimate the change in period for
two choices of a constant $G$: in an idealized, completely unscreened star
the first corresponds to the coupling constant $\alpha_c=1$,
and the second to $\alpha_c=1/3$, which applies to all $f(R)$ models (see above). The modified gravitational constant is denoted $G_1$ and $G_2$ for
the two cases, and the corresponding periods $\Pi_1$ and $\Pi_2$ are:
\begin{eqnarray}
G_1 = 2 \ G_N     &\implies & \Pi_1 = \Pi_N/\sqrt{2} \\
G_2 = 4/3\ G_N &\implies & \Pi_2 = \Pi_N/\sqrt{4/3}
\label{eqn:predictions}
\end{eqnarray}
where the subscript $N$ denotes values in the Newtonian gravity. The shorter
period means that in chameleon gravity, the inferred distance (based on
incorrectly using the GR Period-Luminosity relation) is smaller than the true
distance.

A fully unscreened star is an idealization -- it is important to take into
account the spatial profile of $G$ and the fact it is not altered inside
the screening radius. This will tend to lower the deviation. 
We take this into account and estimate a more realistic value of
$\Delta G$ in the unscreened region by averaging $G(r)$ according to
\begin{equation}
\langle G\rangle = \frac{1}{R} \int_0^R f(r)G(r)\dd r
\end{equation}
where $G(r)$ is given by equation \ref{eq:g(r)}. The function $f(r)$ is a weighting function that accounts for the fact that different regions of the star are more important than others in driving the pulsations. The simplest scenario is to simply take $f(r)=1$, however Epstein (1950) has numerically solved the pulsation equation and tabulated the weight function $f$ (see figures 1 and 2 in his paper). Using the numerical values given in the tables, we have reconstructed the normalised weight function.
We use the Epstein function in conjunction with $G(r)$ profiles from MESA along the instability strip to obtain $\langle G\rangle$. Figure \ref{fig:G-profiles} shows the 
actual profiles $G(r)$ for different choices of $\alpha_c$ and $\chi_c$. 

To estimate the change in distance, we need the Period-Luminosity relation for a given observational band.
If one uses $\rho \sim M/R^3$ in Equation \ref{eqn:period1}, and $L = 4\pi
R^2 \sigma T_e^4$,
 then one gets a relation between $\Pi$, $L$ and $T_e$ that is nearly universal
for all cepheids.
 The main residual dependence is on metallicity. Using observational quantities,
such as
 the V-band absolute magnitude $M_V$ (note though
 that it is the bolometric magnitude $M_B$ that is
 directly related to $L$) and (intrinsic) color $B-V \propto {\rm log}\ T_e$
gives
 \begin{equation}
 M_V = \tilde\alpha\ {\rm  log}\ \Pi + \tilde\beta (B-V) + \tilde\gamma
 \label{eqn:PLC}
 \end{equation}
 where $\tilde\alpha$ and $\tilde\beta$ are universal in galaxies with similar metallicity,
e.g. the Milky
 Way and other galaxies dominated by Pop II stars. For the observations discussed below a reasonable approximation is $\tilde\alpha\approx -3$.
 Using the
$P-L$ relation above and the fact that
the flux goes as $L/d^2$, the change in inferred distance is
\begin{equation}
\frac{\Delta d}{d} \approx -0.3 \ \frac{\Delta G}{G}, \ \ \ {\rm where} \ \ \ 
\frac{\Delta G}{G} \equiv \frac{\langle G \rangle - G_N}{G_N}.
\end{equation}
Table 1  gives the effective values of $\Delta G/G$ obtained from MESA
and the resulting change in distance\footnote{The absolute
value of $\tilde\alpha$ is significantly larger in the infrared, which would lead to larger changes in distance. It also provides an additional
check on gravity: the inferred distance should vary with filter. We estimate a
change in distance of about 5\% between the $V$ and $K$ band.}.

\begin{table*}[t!]
\begin{center}
\caption{Change in inferred distance due to the change in cepheid periods for different gravity parameters. For a 6$M_\odot$ cepheid, the change in effective $G$ (using the Epstein weights as described in the text) and inferred distance is shown for different values of the coupling constant $\alpha_c$ and background field value $\chi_c$. All $f(R)$ models correspond to $\alpha_c=1/3$ with $\chi_c\equiv f_{R0}$.    }
\label{table:period}
\bigskip
\begin{tabular}{l  c  c  c }
\hline
$\alpha_c$ & $\chi_c$ & $\Delta G/G$ & $\Delta d/d$ \\
\hline
1/3 & $4\times 10^{-7}$ & 0.11 & -0.03\\
1/3 & $1\times 10^{-6}$ & 0.21 & -0.06\\
1/2 & 4$\times 10^{-7}$ & 0.17 & -0.05\\
1/2 & 1$\times 10^{-6}$ & 0.34 & -0.09\\
1 & $2\times 10^{-7}$ & 0.21 &  -0.06\\
1 & $4\times 10^{-7}$ & 0.45 & -0.12  \\
\hline
\end{tabular}
\end{center}
\end{table*}

Note that we test for gravity using these predicted changes in distance using a sample of galaxies each of which has dozens to hundreds of observed cepheids. While systematic errors in absolute distances are a challenge, as is theoretical uncertainty, 
we use relative distances in our tests, so some gain in accuracy is achieved
via averaging over many cepheids and galaxies.

The above estimates are based on simple approximations for the theory of cepheid pulsations. 
We do not address the amplitude of the
oscillations, which depends on luminosity and other properties.
Nor does our analysis deal with the location of the instability strip which involves the
Luminosity-Mass-effective Temperature relation; the instability strip itself is quite
narrow, $\simeq 200$K. The amplitude and shape of the pulsations, the precise value of the period, as well as location of 
the instability strip are reasonably well understood and well measured. Changes 
in these properties are therefore additional possible
tests of gravity. The 
computation involves the use  
of detailed non-linear, non-adiabatic models that are well studied in the literature 
(though the treatment of 
convection and metallicity dependence remain somewhat uncertain). 
For MG, this also requires modeling the evolution up to the 
instability strip via MESA. 
These issues are beyond the scope of this study and
will be addressed elsewhere.

We use only the distances obtained from the $P-L$ relation for our tests below: 
it is worth noting that our analysis is robust to the effects of MG 
on the luminosity and radius of the star. To see this note that in obtaining 
a distance estimate from the $P-L$ relation of Eqn. \ref{eqn:PLC}, the observables 
are the flux $f$, $T_e$ and $\Pi$ while the coefficients $\tilde\alpha$
and $\tilde\beta$ follow from $L = 4\pi\sigma R^2 T_e^4$. 
So the entire dependence on MG is via $G$; the change in $G$ may be regarded as  contained in the coefficient $\gamma$.  In practice $\tilde\alpha$
and $\tilde\beta$ are set empirically since observations involve filters with finite wavelength coverage, so estimates of the flux and temperature are imperfect. 
We have used the empirical value $\tilde\alpha \approx -3$ to check that the 
residual dependence on $L$ is weak: $\Delta d/d \approx -0.025 \Delta L/L$ at
constant $T_e$ -- at least a factor of ten smaller than the signal from MG models we consider. An additional effect is the mass dependence of MG effects 
(more massive stars  have slightly larger force enhancements). This leads to shallower
predicted slopes of the $P-L$ relation and other second order effects that we
do not consider  here but merit further study. 

\subsection{Observations of Cepheids}

 Cepheid distances have been calibrated using parallaxes for
 10 Milky Way cepheids in the distance range $\approx 0.3-0.6$ kpc, with periods
ranging from $\approx 3-30$ days.
 The error on the mean distance is $\pm 3\%$ or 0.06 magnitude. The slope has in the past been obtained from the LMC since the sample size is bigger, but at the cost of   possible uncertainty due to 
metallicity effects. More recently the maser distance to NGC 4258
 has superseded the calibration with the LMC. Outside the Local Group, 
 cepheid distances have been measured to over 
50 galaxies (see the review by Freedman \& Madore 2010 for more details).
The final uncertainty in the distance modulus, which includes zero point calibration, metallicity, reddening and
other effects, is $\pm 0.09$
magnitude or 5\% in distance.
\footnote{The {\it GAIA} space telescope will improve cepheid calibration. An
important recent development is the measurement of the $P-L-C$ relation in the IR.
The slope is steeper and the scatter is significantly smaller in the IR, so
{\it Spitzer} and JWST should improve the calibration. A factor of two improvement is anticipated -- see Table 2 in Freedman \& Madore (2010). }
As discussed in Sasselov et al (1997) and subsequent work, the three basic ingredients
in the $P-L$ relation (pulsation theory, stellar evolution and stellar atmospheres)
are sensitive to metallicity. A metal-poor cepheid is
fainter for given period and temperature.
The net dependence on metallicity however is weak, in
particular the slope of the relation between period and bolometric luminosity is nearly
unchanged.

\begin{figure}[t]
\centering
\includegraphics[width=12cm]{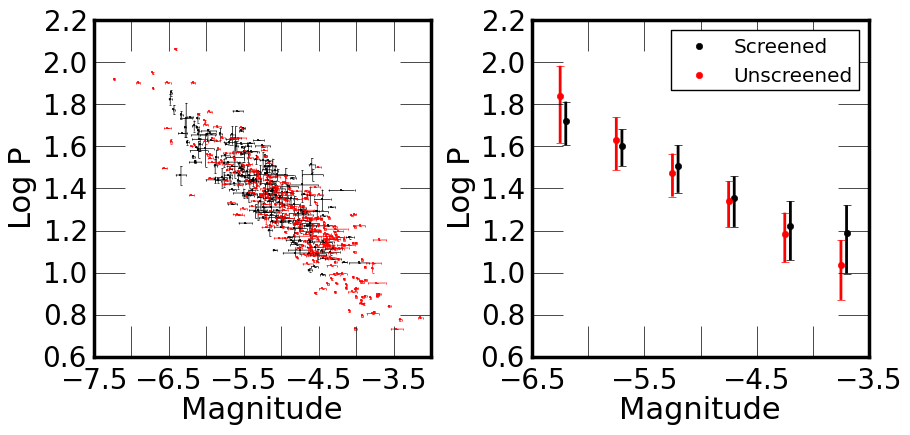}
\caption{The $P-L$ relation for the galaxies in our sample. 
In the left panel, we show all the cepheids observed along with  
the reported error bars. The right panel shows the mean period and dispersion
within bins in absolute magnitude of size 0.5. 
The red and black points show unscreened and screened
galaxies respectively. There is no evidence for a difference in the shape of
the $P-L$ relation between the two samples.
}
\label{fig:cepheids2}
\end{figure}

The data for the $P-L$ relation used in our analysis was compiled for individual  cepheids in 19 galaxies -- see Appendix C for details.  Five galaxies were removed from the sample due to the large scatter in the relation.  Of the remaining 14 galaxies used for Figure \ref{fig:cepheids2}, seven galaxies 
have TRGB based distances as well. 
The majority of the galaxies show late type morphology and include both dwarf and normal galaxies with peak
rotational velocities ranging from 40 to 240 km/s.  
The number of cepheids in each 
galaxy varies between 5 and 117, and is in the range 10-50 for most galaxies.
We use only cepheids with good photometry in V-band. We use the
phase-weighted cepheid luminosity when possible (available for the majority of
the cepheids) and intensity-weighted luminosity for the remainder. 
Compiling this dataset requires a detailed compilation of data on hundreds of  
cepheids, so we have used only a subset of 
existing data.  The sample size can be significantly increased to carry out 
additional tests, given more
detailed theoretical predictions for MG as discussed below. 

We classify these galaxies as screened or unscreened based on their mass (which
determines whether the self-screening condition is satisfied) and
their environment. No rigorous criteria exist to determine the screening effect
of the neighboring galaxies, groups and clusters since the equations are
nonlinear. However, based on recent work
(Zhao et al 2011; Cabre et al 2012) and on tests we have performed, we
 use an estimate of the mean Newtonian potential over the galaxy
 due to its neighbours within a (background) Compton wavelength. This is estimated
 from the SDSS, 2MASS and other surveys, and used
to determine the screening condition in combination with Eqn. \ref{eqn:screening_vmax} for  self-screening 
(details are presented in Cabre et al 2012). 

Figure \ref{fig:cepheids2} shows the observed $P-L$
relation in our sample. We use the normalization 
from Saha et al (2006) 
to find the absolute magnitude. The left panel shows the individual data points 
with error bars and the right panels shows the averaged $P-L$ relation 
in bins of absolute magnitude. The red points show cepheids in unscreened
galaxies and black points in screened galaxies. The slope of the $P-L$ relation
is consistent between the two samples. The unscreened sample has a slightly 
steeper slope -- we have checked that the expected signal from MG is the opposite, based on changes in the period which is shorter for massive cepheids. 
We do not use this signature as a test here since there are uncertainties in the model predictions. However, this is a potential future test for MG. 

\section{TRGB: Tip of the Red Giant Branch Distances}

\begin{table*}[th!]
\begin{center}
\caption{Change in inferred distance using the TRGB peak luminosity for different gravity parameters. For a 1.5$M_\odot$ red giant, the change  in luminosity and the inferred distance is shown for different values of the coupling constant $\alpha_c$ and background field value $\chi_c$. All $f(R)$ models correspond to $\alpha_c=1/3$ with $\chi_c\equiv f_{R0}$.    }
\label{table:period}
\bigskip
\begin{tabular}{l  c c   c } 
\hline
$\alpha_c$ & $\chi_c$ & log $L/L_\odot$ & $\Delta d/d$ \\
\hline
0 & 0 & 3.34 & 0 \\
1/3 & $1\times 10^{-6}$  & 3.32 & 0.02\\
1/3 & $2\times 10^{-6}$  & 3.30 & 0.04\\
1/3 & $4\times 10^{-6}$  &  3.25 & 0.12\\
1/3 & $8\times 10^{-6}$  & $<$3 & $>$0.20\\
\hline
\end{tabular}
\end{center}
\end{table*}

The TRGB distance is obtained by comparing the measured flux to the 
universal peak luminosity expected for red giants with masses below 2 $M_\odot$. 
TRGB distances have been measured to $\sim 250$ galaxies using the universality
of the peak luminosity of red giants at the tip of the red giant branch. These
are applied to old, metal poor populations which enables distance estimates
out to about 20 Mpc, a bit closer than cepheid distances since cepheids are
brighter. However since single epoch photometry suffices, it is much easier
to obtain the data for a TRGB distance. While it is not an absolute distance
method, and must in fact be calibrated using secondary indicators like
cepheids, a comparison of TRGB distances in screened and unscreened
galaxies can still provide a useful test.

The TRGB luminosity is set by the He flash during the
post main sequence evolution. Low mass stars develop a small He core region
after the main sequence (which grows slower than that of high mass stars).
In the initial stage of post-main sequence evolution the core temperature is
not high enough to ignite the He and the core subsequently contracts due to the absence of outward pressure. This catalyses the nuclear reaction in the
outer hydrogen dominated region and the He produced in the shell gets
deposited on to the core. The increase in the mass of the core
causes the shell luminosity to grow -- for stars of interest the
luminosity of the star is roughly $\propto M_c^{23/3}/R_c^6$, where $M_c$ and
$R_c$ are the mass the radius of the core. When the core temperature becomes high enough ($T_c \sim 10^8K$) to ignite
He, the star moves to the left in the H-R diagram, settling onto the horizontal branch. Nuclear physics within the core sets
the TRGB luminosity, mostly independent of  composition or mass of the envelope. 

The luminosity of the TRGB depends on the mass of the core which in turn depends
weakly on the metallicity of the star. In particular, for stars with low
metallicities ($-2.2 \leq $ [Fe/H] $ \leq -0.7$) the $I$ band magnitude of the
TRGB varies by only 0.1 magnitude within this metallicity range ($I_{TRGB} = -4.0 \pm 0.1$), though it can vary a lot more in other bands.
Figure 1 in Lee, Freeman \& Madore (1993) shows the
variation of $M_V$ and $M_I$ over the above metallicity range. The remarkable
constancy of the TRGB magnitude leads to a discontinuity in the luminosity function
of stars as low mass stars continuously reach this phase. The distance to the
TRGB can be measured if we can filter out the magnitude at the discontinuity.
Observationally, the TRGB is identified from a semi-empirically calibrated
color-magnitude diagram (Da Costa \& Armandroff 1990) and the metallicity
corrected distance modulus is determined.

We have estimated the change in luminosity with modified gravity for the TRGB
using MESA (see Appendix B for a discussion and an analytic estimate). Figure 
\ref{fig:G-profiles} shows the profiles of $G(r)$ at the TRGB stage for a 1.5$M_\odot$ star for different values of $\chi_c$ with $\alpha_c=1/3$. There is a transition in the enhancement of $G$ in the shell as $\chi_c$ increases from $10^{-6}$
to $10^{-5}$. 
We find that for $\alpha=1/3$, the resulting 
change in luminosity is at the percent level for $\chi_c < 10^{-6}$.
However for $\chi_c$ between $5\times 10^{-6} - 10^{-5}$, the shell becomes unscreened and the luminosity changes rapidly, falling by $20\%$ to over $50\%$ in this range. The change in distance that would result is detectable -- a fact used in the comparison of TRGB and maser distances below. Table 2 gives the change in inferred distance $\Delta d/d$ for different values of $\chi_c$.




The TRGB data used here is taken from the literature and involves many
telescopes, including  WFPC2/ACS on
board the Hubble Space Telescope, the 5m Hale telescope at Palomar Observatory,
Isaac Newton Telescope Wide Field Camera, Wisconsin Indiana Yale NOAO 3.5m
telescope, VLT etc. The photometry on these images was done using the DAOPHOT
and/or ALLFRAME (Stetson 1987, 1994) packages. The observed magnitudes are
corrected for foreground extinction. Methods like Sobel edge-detection (Lee,
Freeman \& Madore 1993), maximum likelihood (M\'{e}ndez et al. 2002) etc. were
used to estimate the TRGB magnitude. We classify the galaxies into screened and
unscreened samples as for the cepheids.

\section{Results: Constraints on Gravity Theories with Cepheid and TRGB Distances}

\begin{figure}[t,h]
\centering
\includegraphics[width=16cm]{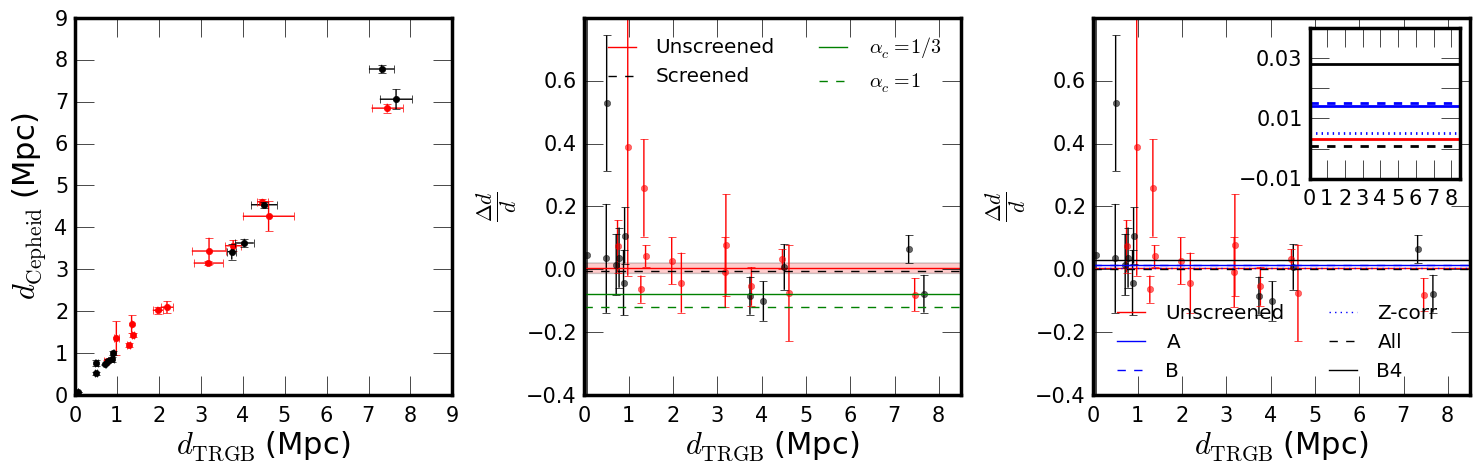}
\caption{{\it Left panel:} A comparison of distances measured using 
the cepheid $P-L$ relation and TRGB
luminosities. The black points are for
screened galaxies and the red points for unscreened galaxies.
 {\it Middle panel:} $\Delta d/d$, the fractional
 difference between cepheid and TRGB distances, as a function of TRGB distance. 
 The shaded
region in the middle panel shows the 68\% confidence region around our best
fit to the unscreened sample (red line). The best fit to the screened sample is
shown by the dashed black line. The data are consistent with the GR expectation of
zero deviation in distance and there is no visible
trend in the deviation with whether the galaxy is screened or not. 
The dotted and dashed green lines show two possible predictions of chameleon theories with 
$\alpha_c = 1$, $\chi_c = 4 \times 10^{-7}$ and $\alpha_c = 1/3$,
$\chi_c = 1 \times 10^{-6}$ which corresponds to $f(R)$ gravity. 
{\it Right panel:} As checks on systematic errors and model uncertainties, the four 
 lines show the best fits obtained with other choices for the screening threshold or metallicity correction -- see text for details. These lines lie within the 68\% confidence region shown in the middle panel. 
}
\label{fig:comparison}
\end{figure}

We began with a sample of 27 galaxies with both TRGB and cepheid distances
taken from the literature compiled by Madore \& Steer using 
NED\footnote{http://ned.ipac.caltech.edu/level5/NED1D/ned1d.html}. 
This sample includes
12 galaxies from the Hubble Space Telescope Key Project (Freedman et al. 2001).
Other galaxies in their sample do not have TRGB distance measurements. For many
galaxies more than one measurement exists both for TRGB and cepheid based
distances. We use bootstrap resampling to obtain the average and error bars on 
those measurements. We exclude one galaxy with TRGB distance beyond 10 Mpc and another galaxy (DDO 187) which has only two confirmed cepheids. 
This leaves us with 25 viable galaxies. 

We perform a likelihood analysis on the data to estimate the best fit value of $\Delta
d/d$. The best values and 1-$\sigma$ errors are given in Table 3 along with the reduced $\chi^2$.  We included  empirically estimated systematic errors in the estimate of the distance to each galaxy from multiple measurements, as well as in the average deviation $\Delta d/d$ for each subsample of galaxies. For the latter we made the ansatz that each galaxy has an additional unknown systematic error that can be added in quadrature to the reported error. By further assuming that the systematic error was the same for each galaxy, we could estimate $\sigma_{\rm sys}$ iteratively such that the reduced $\chi^2 $ was unity. We found that the systematic error thus
estimated is subdominant for the majority of galaxies. 

We have tested our estimate of statistical and systematic errors by using 
several different methods of weighting the multiple distance measurements 
for each galaxy, of outlier rejection and bootstrap resampling.  The 
various estimates lie within the 1-$\sigma$ interval shown and generally tend to 
deviate towards positive values (i.e. further away from the MG predictions; see below). 
Note that for higher values of $\chi_c$ the TRGB distances are also affected in a way that increases $\Delta d$: the inferred distance would be larger in modified gravity since the peak luminosity is lower. So the
predicted deviation from cepheid distances increases for $\chi_c \gsim 10^{-6}$.

\begin{table*}[t!]
\begin{center}
\caption{Best fit values for $\Delta d/d$ and uncertainty $\sigma$ in  the fractional  difference between
cepheid and TRGB distances, for screened and unscreened subsamples. Our estimated $\sigma$ includes systematic errors; the number of galaxies $N$ and 
reduced $\chi^2$ is also given. 
}
\label{table:period}
\bigskip
\begin{tabular}{l  c  c  c c }
\hline
Sample & $N$ & $\Delta d/d$ & $\sigma$ & Reduced $\chi^2$\\
\hline
Unscreened & 13 & 0.003 & 0.017 & 1.0\\
Screened & 12 & -0.005 & 0.022 & 1.3\\
\hline
\end{tabular}
\end{center}
\end{table*}

 Figure \ref{fig:comparison} shows the cepheid distance vs. the
TRGB distance for this sample of galaxies, separated into screened (black points) and
unscreened (red points) subsamples.
The middle and right panels show $\Delta d/d$, the fractional difference
between the cepheid and TRGB distances. 
The shaded region in the middle panel 
shows the 68\% confidence region around our best
fit (red line) for the unscreened sample. The dashed black  line shows the best fit for the screened sample -- it nearly overlaps the red line. 
Clearly the screened and unscreened samples are consistent. 
The two green lines show the predictions 
for chameleon theories with coupling strength $\alpha_c=1/3, 1$ and values of $\chi_c$ as indicated. The two models shown are ruled out at over 95\% confidence. 

In the right panel of Figure \ref{fig:comparison} we show some alternate 
estimates of $\Delta d/d$. 
Since our estimate of the Newtonian potential due
to neighbours is subject to uncertainties in the galaxy catalogue, we
show two other  screening criteria, labelled A and B, with threshold 
$\Phi_{\rm N} = 2\times 10^{-7}$ and 
$1\times 10^{-6}$ (our fiducial choice is $4\times 10^{-7}$). We also show 
the result obtained using the full sample of galaxies (labelled All). 
The three best fits are within the 68\% 
confidence region of the middle panel, in fact they deviate in the opposite direction from the MG model predictions. This indicates that variations in the 
screening criterion, as required for other choices of chameleon theory parameters or for symmetron screening, do not weaken our result. We also show the best 
fit obtained using a simple average over the best 4 measurements per galaxy (B4). 
Finally the best fit value using the metallicity 
correction of Sakai et al (2004) is also shown (Z-corr). The majority 
of the distance estimates we used did not attempt such a correction. 
We have not used it in our 
fiducial best fit since there is a slight correlation of metallicity
with level of screening which may introduce correlations with the
signal; moreover the goodness of fit 
for our sample did not improve with the Sakai et al metallicity
correction.  With a large enough sample of galaxies, one can 
attempt to create screened and unscreened subsamples that have similar
metallicity distributions and thus do a controlled metallicity
correction.  

We note that cepheid and TRGB distances are calibrated using cepheids with 
parallaxes and TRGB stars in globular clusters  in the Milky Way. Thus we rely on the Milky Way being screened in using them as tests of gravity in unscreened galaxies. This means that, as for other astrophysical/cosmological tests, the constraints we obtain for large values of $\chi_c$  (above about $10^{-6}$) require an unconventional interpretation of screening to satisfy solar system tests of gravity, such as a significant effect of the mass distribution of the Local Group. On the other hand, if somehow say TRGB stars in the Milky Way were unscreened, thus implying that the TRGB distances in our sample were calibrated with an unscreened luminosity relation, then there would be strong deviations with stellar mass and host galaxy environment (large groups or clusters) that are likely ruled out by current data. 

\subsection{Constraints on chameleon theories}
Figure \ref{fig:confidence-regions} summarizes our constraints on
chameleon theories. In the $\alpha_c$-$\chi_c$ plane, we plot the
regions excluded at 68\% and 95\% confidence with the light and dark
shaded regions. We have made some approximations in obtaining this
plot, especially in our criteria for screened and unscreened galaxies:
we use a fiducial choice $\alpha_c=1/3$ and vary the subsamples as 
$\chi_c$ varies. The relatively small number of galaxies available is
responsible for the jaggedness in the contours.  
We assume that the environmental screening criterion 
is not sensitive to the value of
 $\alpha_c$ directly. While $\alpha_c$ is expected to change the thickness of the 
 shell around screened objects, observable stars are located well inside
 the galaxy halo.  For our test,  the best fit 
 is very robust to the choice of the unscreened sample, 
 as evident from the right panel of Figure \ref{fig:comparison}: essentially we 
 cannot find any selection of galaxy subsamples that correlates with screening and  produces a statistically significant deviation from GR. 
 
We have also tested the results shown in Figure \ref{fig:confidence-regions} with different screening criteria and several 
methods of estimating systematic errors as described above: 
the 95\% confidence contour is 
robust to all our tests, while the 68\%  confidence contour can vary somewhat. 
A larger galaxy sample and detailed theoretical calculations are 
required to obtain a more rigorous version of
Figure \ref{fig:confidence-regions}. 

We summarize our limits for two specific choices of $\alpha_c$: 
\begin{itemize}
\item $\alpha_c=1/3$: Upper limit at 95\% confidence: $\chi_c \approx
  5\times 10^{-7}$ 
\item $\alpha_c=1$: Upper limit at 95\% confidence: $\chi_c \approx
  1\times 10^{-7}$. 
\end{itemize}
These limits correspond to a background Compton wavelength of $\sim$ 1 Mpc 
for the models discussed in the literature (e.g. Schmidt, Vikhlinin and Hu 2009). 
The very short range of the scalar force 
implies an extremely limited modification of gravity, i.e. a  fine tuned
model.  The limits on both parameters can be extended further in the near future 
with additional analysis and forthcoming data on cepheids. 

Limits from cosmological tests on the background field value are over 
two orders of magnitude weaker than the limits obtained here. 
The cosmological analysis of
$f(R)$ gravity (including SN + CMB + ISW + cluster data) done by 
Schmidt et al (2010) and Lombriser et al (2010)  
gives the upper limit $\chi_c \approx 10^{-4}$ for $\alpha_c=1/3$. This limit
is also indicated in Figure \ref{fig:confidence-regions}. The 
constraining power comes mostly from galaxy cluster counts. The 
constraints on gravity by Reyes et al (2010) that use the test proposed
by Zhang et al (2007) do not constrain $\chi_c\sim 10^{-4}$ even at 68\% 
confidence. Cosmological tests have not been used probe values of 
$\alpha_c$ other than 1/3. While our local astrophysical 
tests are more powerful for chameleon theories,  
it is worth noting that more generally any probe of gravity in a distinct regime of length 
scale and redshift is valuable -- in that sense the local tests are complementary 
to cosmological tests. 

\subsection{Constraints on Symmetron Theories}

The symmetron screening model has some qualitative similarities 
to chameleon screening (see Appendix A). 
Hinterbichler and Khoury (2010) showed that 
solar system tests place constraints on parameters of symmetron 
cosmology that are analogous to $\alpha_c$ and $\chi_c$. Clampitt, 
Jain \& Khoury (2011) computed the screening profile of galaxy halos
of different masses. To translate our results to symmetron models, we
use the following relation of our parameters to those of Clampitt et al: $\chi_c \equiv 1/2 (M_s/M_{\rm Planck})^2$ and $\alpha_c = 2 g^2$. Solar
system tests set the constraint $\chi_c \lsim 10^{-6}$ or $M_s \lsim 10^{-3} M_{\rm Planck}$ for 
$g=1$ and a Compton wavelength of $\sim 1$ Mpc (Hinterbichler \& Khoury 2010). 

Our upper limit for the fiducial symmetron model described above is  
$\chi_c \lsim 3\times 10^{-7}$. We can extend our 
results for other values of $\alpha_c$ to the symmetron parameters as well. We do
not pursue a more detailed analysis here as 
it would require the screening condition to be worked out carefully for the symmetron
case. We do use the mapping from chameleon to symmetron self-screening
described in Clampitt, Jain \& Khoury (2012) but the environmental
screening  needs to be considered more carefully (Joseph Clampitt, private communication).  

\subsection{Additional Tests of Gravity with Cepheids}

We have not considered some additional gravity tests that are possible with distance indicators. These include the following.

\begin{itemize}
\item The location of the instability strip and other properties of cepheids (size, luminosity and pulsation amplitude) are affected by modified gravity. With more detailed theoretical predictions, these provide additional tests (Sakstein et al., in preparation).
\item Variation of the slope of the $P-L$ relation and its dependence on filter. 
Since the periods of  massive cepheids are affected more strongly, the 
$P-L$ relation should have a shallower slope for unscreened galaxies. A second
effect arises from the steeper slope of  the $P-L$ relation
in the IR -- this means that  the inferred distance would be smaller in the IR. 
We see no hints of a signal with the limited available data, but a useful tests requires significantly more observations with Spitzer and other IR instruments.
\item The variation of estimated distance with cepheid mass and temperature and with the degree of screening of the host galaxy. These would require 
detailed theoretical predictions, a high resolution screening map for different galaxies
and a far more detailed analysis of observations than we have performed. 
\end{itemize}



\section{Masers and Other Distance Indicators}

The comparison of distances from cepheids or TRGB with other methods that rely
on self-screened objects can provide useful tests. Distances obtained using Type Ia Supernovae (SN) are likely unaffected by MG for $\chi_c \lsim 10^{-4}$, while  maser
distances use a purely geometric method so they are unaffected by MG. 
We do not attempt to create a screened vs.
unscreened galaxy sample in this section 
since maser and SN distances are not available for
sufficient numbers of dwarf galaxies. We rely on the calibration of cepheid or TRGB
distances in the Milky Way (taken to be screened either by its own potential or that of the Local Group) and work in the parameter range where these distances
are affected by modified gravity. As discussed above, along 
with other astrophysical or cosmological tests, 
the logic of pursuing constraints at field values $\chi_c > 10^{-6}$ 
requires invoking an unconventional source of screening for the solar system. 

\subsection{Water Masers and the Distance to NGC 4258}

Maser distances are inferred by comparing the rotation velocity
and proper motion (angular velocity) of water masers in Keplerian motion around
supermassive black holes in spiral galaxies. Measurement of the centripetal acceleration provides a second distance estimate.
NGC 4258 is a Milky Way sized galaxy at a distance of 7 Mpc.
The water masers in this galaxy provide a rotation velocity of the accretion disk 
 in excess of 1,000 km/s at distances on the order of
0.1-0.3 pc from the inferred super-massive black hole of mass
$4\times 10^7 M_\odot$. 
 The two distance estimates obtained
from the maser data  are in excellent agreement --
see Herrnstein et al (2005) and Humphreys et al (2008)
for recent studies.

A summary of the maser, cepheid and TRGB distance estimates to NGC 4258 following Freedman \& Madore (2010) is:
\begin{eqnarray}
{\rm NGC 4258\ Maser}: d &=& 7.2 \pm 0.2\ {\rm Mpc} \\
{\rm NGC 4258\ Maser}: d &=& 7.1 \pm 0.2\ {\rm Mpc} \\
{\rm NGC 4258\ Cepheid}: d &=& 7.18 \pm 0.07 ({
\rm statistical})\ {\rm Mpc} \\
{\rm NGC 4258\ TRGB}: d &=& 7.18 \pm 0.13 \pm 0.40\ {\rm Mpc}
\end{eqnarray}
There are several caveats to the above tabulation, especially for the cepheid
distance which does not include systematic errors, but a full discussion  
is beyond the scope of this
paper (see Benedict et al 2007; di Benedetto et al 2008 and Mager, Madore \& Freedman 2008 for recent cepheid distance estimates). We note that the distances agree within estimated errors that are at the
few percent level for the maser distances and (allowing for systematics) at the 5-10\% level for cepheid and TRGB distances.

The agreement of TRGB and water
maser distances probe field values $\chi_c$ above $10^{-6}$ -- the precise 
range probed depends on 
the value of $\alpha_c$ and the typical mass of the star. 
Specifically for $\alpha=1/3, \chi_c > 4\times 10^{-6}$ and a typical 
star of mass $1.5 M_\odot$ the TRGB luminosity is smaller by over
20\%, corresponding to an inferred distance that is larger by over 10\%.
Thus given the measurements above, $f(R)$ models with this parameter range are excluded. Note that the maser distance is purely geometric: it is a ratio of velocities or a combination of velocity and accelerations that is independent of the strength of gravity; it is immune to MG effects. Thus higher field values up to the cosmological upper limit of $10^{-4}$ are also excluded since TRGB distances would change drastically for these values.

The agreement with cepheid distances would probe lower field values
but NGC 4258 is a Milky Way sized galaxy, so for lower
field values the galaxy would screen the cepheids.
One must wait for future observations of
masers in smaller galaxies to probe field values below $10^{-6}$ though
it is unclear whether useful maser distances can be found for galaxies with much 
smaller Newtonian potentials. 

\subsection{Other Distance Indicators}

Type Ia SN are a valuable distance indicator at cosmological distances
where cepheid and TRGB methods cannot be applied. 
The energetics of a Type Ia SN is set by the 
thermonuclear fusion of Carbon-Oxygen nuclei in the core of
the progenitor white dwarf. The emission responsible for the
observed optical light curve comes from
the expanding shell in which Nickel nuclei decay into Iron and release high
energy photons that subsequently thermalize. While the Newtonian
potential at the surface of the core is $\sim 10^{-4}$, it is smaller at the
distance of the expanding shell. We have not attempted to use supernovae distances as a test here because SN distances are calibrated using cepheid distances for 
a small number of galaxies with both distance indicators available. With a much larger sample, using a subset of screened galaxies for calibration may make an 
unscreened subsample available for tests. 

Finally, other widely used distance indicators include Tully-Fisher,
Fundamental Plane and
Surface Brightness Fluctuations. The first two methods rely on the dynamics of
stars in disk and elliptical galaxies. The scatter in these methods remains
large and their use would require averaging over large numbers of galaxies.
We do not pursue these methods here.

\begin{figure}[t]
\centering
\includegraphics[width=10cm]{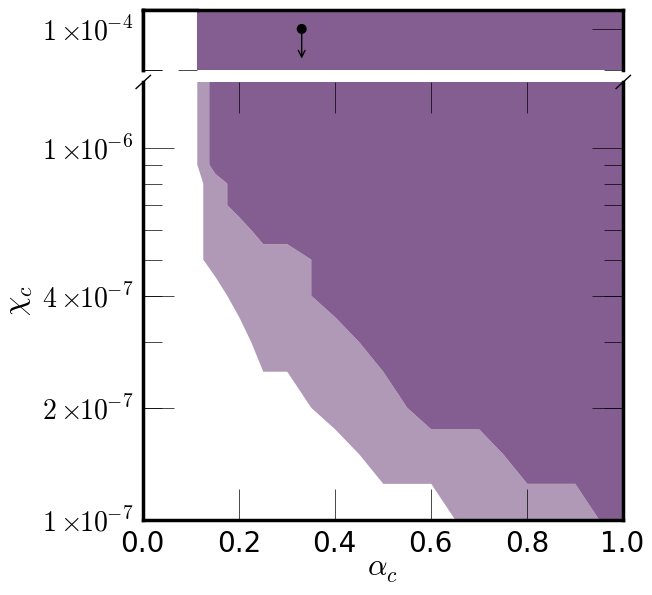}
\caption{Upper limits on the two parameters of chameleon theories:
 the coupling parameter $\alpha_c$ and the background field value $\chi_c$. 
 The boundaries of the shaded regions show the upper
 limits at 68\% and 95\% 
confidence level. These are obtained using an interpolation of
our tests for the two gravity parameters as discussed in the text.  The effects of 
discreteness are due to the small sample of galaxies used. The upper 
end of the $y$-axis is extended to $\chi_c=10^{-4}$ to show the upper limit from
cosmological+cluster constraints which was obtained for the $f(R)$ model parameter $f_{R0}\equiv \chi_c$ with $\alpha_c=1/3$. 
}
\label{fig:confidence-regions}
\end{figure}

\section{Discussion}

We have used low-redshift distance indicators
to carry out tests of scalar-tensor gravity theories. In particular, since
different distance indicators operate in gravitational
fields of different strengths, their screening behavior varies.
A comparison of  distance estimates from cepheids, TRGB stars and other
distance indicators in nearby dwarf galaxies can provide 
powerful tests of gravity. The results shown here are applicable to chameleon
theories (including all $f(R)$ models) and we have shown rough constraints on 
 symmetron screening as well. Indeed, with the notable exception of theories 
that use Vainshtein screening, a generic scalar-tensor gravity theory is likely to be 
constrained by our analysis. Our tests also constrain  
scalar field couplings to matter that may arise in dark energy, string theory or other scenarios.

Cepheid variables are the least compact of the distance indicators
considered here -- the amplitude of the surface Newtonian potential  is typically 
 $\Phi_{\rm N} \sim 10^{-7}$ (for TRGB stars the relevant 
 $\Phi_{\rm N} \gsim 10^{-6}$). 
 Thus in unscreened galaxies, cepheids may
experience enhanced forces due to the scalar field
-- this will lower the inferred cepheid
distance compared to screened distance indicators. 
It turns out that the deviation of the inferred distance for cepheids and TRGB 
 stars is in opposite directions, so for  field values in which both are 
 unscreened they would 
 show even larger discrepancies. 
 
We have shown that current data is consistent with GR and is inconsistent with 
chameleon theories over a parameter range that is more than 
two orders of magnitude below 
previous astrophysical tests. Figures \ref{fig:comparison} and \ref{fig:confidence-regions} show our upper limits for the two parameters: the coupling $\alpha_c$ 
and the background field value $\chi_c$. 
For chameleon theories with  
$\alpha_c=1/3$ (all $f(R)$ models) the upper limit on   
$\chi_c$ is about $5\times10^{-7}$ at 95\% confidence. 
We show results for values of $\alpha_c$ in the range $0.1-1$. The upper limit 
on $\chi_c$ drops just below $10^{-7}$ for $\alpha_c=1$. 
The comparison of maser and TRGB distances
to NGC 4258 provides an independent test of field values $\chi_c > 2\times 10^{-6}$.

Cosmological observations so far have probed field values larger than $10^{-4}$ 
(Reyes et al 2010; Schmidt, Vikhlinin \& Hu 2009; Lombriser et al 2010 and references therein). Thus our limits exceed the combined analysis of cosmological probes by over two orders of magnitude.  Our upper limits also exceed
solar system and lab tests for some range of chameleon potentials 
(see e.g. the  discussion in Hu \& Sawicki 2007 on the comparison of 
field values in galaxies vs. local tests). With better data on 
distance indicators, lower values
of $\alpha_c$ and $\chi_c$ can be tested, though going much below 
$\alpha_c\approx 0.1$ or $\chi_c\approx 10^{-7}$ will be difficult (due to systematic errors and the self-screening of cepheids, respectively).  
Observations of nearby dwarf galaxies may probe lower field values, which we explore 
in a separate study (Vikram et al, 2012). We note that 
theoretical considerations have also been
shown  to limit acceptable chameleon theories by treating them as effective theories and requiring small quantum corrections (Upadhey, Hu \& Khoury 2012; see also 
Faulker et al 2007). 


Thus the primary advantages of local astrophysical tests such as ours are as follows. 1. The signal is
stronger -- it can be a large fraction of the maximum force modification due to the scalar field. In contrast, the effect on the growth of cosmological fluctuations is typically  
much smaller. 2. Constraints are 
more general and translate  directly to the two key parameters of the theories.  
3. The availability
of a control (screened) sample enables robustness to several systematics. 
4. There is almost no degeneracy with other cosmological parameters or assumptions. 
The primary disadvantage is the presence of astrophysical uncertainties: 
metallicity, galaxy age and stellar population, extinction and so on. However 
many of these systematics do not affect our tests at lowest order since all our tests 
are relative, as discussed above (this allows us to carry out tighter tests than say the distance ladder  which requires an absolute calibration). 

Nevertheless there are three significant sources of uncertainty in our analysis:
an incomplete modeling of the theory, systematic errors in the data, and
approximations used in determining the screening level of the host galaxies.
We have  used results reported in the literature on high quality distance measurements, as summarized in Freedman \& Madore (2010). We attempted to use multiple  weightings of the data as well as  different estimates of systematic errors
to test for the robustness of our conclusions.
Even so, we note that the data is inhomogeneous and our understanding of the
underlying systematics is limited. Ideally by starting with data on individual 
cepheids a more careful and complete analysis can be performed. 

There are a number of open questions for theoretical work that can sharpen the tests reported here and enable new tests. Predictions for the screening level and effective $G$ in symmetron/dilaton screening scenarios would enable these theories to be tested in detail with the same datasets. Non-adiabatic numerical models of cepheid pulsations are needed to improve on the approximate predictions made here. 
We have recently learnt that the predicted deviations of the period that we have 
used are close to estimates with such numerical models 
(which in fact appear to be slightly higher -- 
P. Chang and H. Saio, private communication). Several 
additional tests can be 
carried out with complete theoretical models that incorporate MG, 
as discussed in \S 5.3.

These observational phenomena have been known for a long time -- here we have demonstrated a new use for them as tools for testing gravity. Future observations designed with this in mind could obtain more powerful constraints on these theories. 
Cepheid and TRGB distances to dwarf galaxies out to about 20 Mpc, in a variety of screening environments, are needed to check against relative systematics and
improve constraints on gravity. In this respect it would be prudent to carry out new analyses that are designed to be immune to ``confirmation bias".
Infrared observations of the cepheid $P-L$ relation can provide a strong new test
through the variation of slope of the $P-L$ relation as discussed above. Currently, there is only one galaxy with simultaneous cepheid, TRGB and maser measurements. Maser distances to additional galaxies, especially lower mass galaxies, would strengthen the test described in \S 6.1. Tests that compare SN and cepheid
distances are also worth pursuing. 


%

\bigskip

{\it Acknowledgements:} We are extremely grateful to Bill Paxton for developing and making public the software package MESA and answering our numerous questions. We are very grateful to Anna Cabre, Joseph Clampitt, Anne Davis, Lam Hui, Mike Jarvis, Eugene Lim, Hideyuki Saio and especially Philip Chang for discussions and related collaborative studies. We acknowledge helpful discussions with Gary Bernstein, Wayne Hu, Justin Khoury, Kazuya Koyama, Adam Lidz, Raquel Ribeiro, Abhi Saha, Fabian Schmidt, Mark Trodden, Amol Upadhye, Jake VanderPlas and Gongbo Zhao. BJ is partially supported by NSF grant AST-0908027. JS is supported by the STFC.  

\appendix

\section{Scalar-Tensor Modifications of Gravity with Screening mechanisms}


The density dependent screening mechanisms that we constrain in this work can arise through the scalar-tensor action
\begin{equation}\label{eq:stact}
 S=\int\dd^4x\sqrt{g}\left[\frac{\mpl^2}{2}R-\frac{1}{2}\nabla_\mu\phi\nabla^\mu\phi-V(\phi)\right]+S_{\rm m}[\Psi_{\rm i};\tilde{g}_{\mu\nu}],
\end{equation}
where $S_{\rm m}$ denotes the action for the matter fields and $\Psi_{\rm i}$ represent all matter species in the system. This action looks a lot like the usual action for GR except that the matter fields are not coupled to gravity via the metric but instead via the conformally scaled, \textit{Jordan Frame} metric $\tilde{g}_{\mu\nu}=A^2(\phi)g_{\mu\nu}$. Here $A(\phi)$ is an arbitrary function of a new scalar $\phi$ which is known as the \textit{coupling function}. This action is known as the \textit{Einstein frame} action since the scalar field $\phi$ is coupled non-minimally to the Ricci scalar. One could instead conformally transform to the Jordan frame where the fields couple minimally to the metric but the scalar itself is non-minimally coupled to gravity via the Ricci scalar $A^{-2}(\phi) R$. In this work we shall work exclusively in the Einstein frame since this is where all of our physical tests of GR lie. In this frame, matter fields follow geodesics of the $\tilde{g}_{\mu\nu}$ whilst observers in Einstein frame travel along geodesics of $g_{\mu\nu}$ and so these observers see an additional or \textit{fifth} force given (per unit mass) by (see Waterhouse 2006)
\begin{equation}\label{eq:5force}
 F_{\phi}=\frac{\beta(\phi)}{\mpl}\nabla\phi\quad \mathrm{where}\quad \beta(\phi)\equiv\mpl\frac{\dd \ln A}{\dd \phi}
\end{equation}
 is known as the \textit{coupling}. There are two methods by which the scalar-tensor screening mechanism can act. Either the mass of the scalar is very large (the field gradient is small) so that the force is Yukawa suppressed or the coupling $\beta(\phi)$ is very small and the force is negligible. The first mechanism is utilised by the chameleon mechanism (Khoury \& Weltman 2004), whereas the second is utilised by the symmetron (Hinterbichler \& Khoury 2010) and the environmentally dependent dilaton (Brax et al. 2010) 
 
In this section we shall first describe how these conditions can be achieved in general in the neighbourhood of our solar system before elucidating this further with some common examples.

\subsection{The General Mechanism}

Varying the action with respect to the field $\phi$ results in the equation of motion
\begin{equation}\label{eq:eqmot}
 \Box\phi=V_{,\phi}-T_{\rm m}(\ln A)_{,\phi}
\end{equation}
where $T_{\rm}$ is the trace of the energy-momentum tensor in the Einstein frame. In standard GR we have $T_{\rm m}=-\rho$ where $\rho$ is the matter density, however, due to the non-minimal coupling to the scalar field it is the Jordan frame and not the Einstein frame density that is covariantly conserved. It can be shown (Waterhouse 2006) that the quantity $A^3\rho$ satisfies the non-relativistic continuity equation and so, taking this as our matter density from here on, equation \ref{eq:eqmot} defines a density dependent effective potential for $\phi$
\begin{equation}
 V_{\rm eff}(\phi)=V(\phi)+\rho A(\phi).
\end{equation}
It is this density dependence of the effective potential that is responsible for the screening mechanism. Consider a body of high density (e.g. a star or galaxy) immersed inside a much larger medium of smaller density (the universe) and suppose that the effective potential has a minimum. The minimum of the effective potential will lie at different field values depending on the density and so the field will try to minimise this potential inside both media. If the outer medium is much larger than the high density one then the field will always reach its minimum at asymptotically large distances. Theories with screening mechanisms have the property that either the coupling $\beta$ at the minimum becomes negligible at high densities or the mass of oscillations about said minimum is very large. Hence, if the field can minimise its effective potential effectively inside the high density body then the fifth force will be screened, if this is not possible then the body will be unscreened and the Newtonian force law will receive order one corrections.

To see this more quantitatively, consider a spherically symmetric body of density $\rho_b$ immersed in a much larger medium with density $\rho_c$ with $\rho_b\gg \rho_c$. Inside the medium, far away from the body, the field minimises its effective potential at field value $\phi_c$ and inside the body the field may or may not reach its value which minimises the effective potential, $\phi_b$. If the object is static i.e. on changes occur on a time-scale much smaller than the time-scale for cosmological evolution of the field then equation \ref{eq:eqmot} becomes
\begin{equation}\label{eq:staticphi}
 \nabla^2\phi=V(\phi)_{,\phi}+\rho A(\phi)_{, \phi}.
\end{equation}
 Suppose that the field has reached $\phi_b$ at $r=0$. In this case we have $V(\phi)_{,\phi}\approx-\rho A(\phi)_{, \phi}$ so that equation \ref{eq:staticphi} has no source term and $\phi\approx\phi_b$. In this case there is no field gradient and the fifth force is not present. Of course the field must asymptote to $\phi_c$ away from the body and so there must be some radius $r_{\rm s}$ at which this approximation no longer holds and the source terms become important. In this region the field will posses some gradient and approach its asymptotic values and so we expect order one fifth forces. $r_{\rm s}$ is hence known as the \textit{screening radius}; the region interior to this is screened whilst the exterior region is unscreened. In the unscreened region, the field is a small perturbation about its cosmological value and so we can linearise equation \ref{eq:staticphi} by setting $\phi=\phi_c+\delta\phi$ to find
\begin{equation}
 \nabla^2\delta\phi\approx m_c^2\delta\phi+\frac{\beta(\phi_c)}{\mpl}\delta\rho
\end{equation}
where $\delta\rho=\rho_b-\rho_c$ and $m_c^2=\dd^2 V(\phi)/\dd \phi^2$ is the mass of the field in the cosmological vacuum. On length scales $R\ll1/m_c$ i.e. on those far less than the range of the fifth force in vacuum, we can neglect the first term and the second term can be related to the Newtonian potential via the Poisson equation $\nabla^2\Phi_{\rm N}=4\pi G\delta\rho$. Integrating this twice gives the field in the unscreened region
\begin{equation}\label{eq:phiunscreened}
 \delta\phi(r)\approx-\phi_c+2\beta_c\mpl\left[\Phi_{\rm N}(r)-\Phi_{\rm N}(r_{\rm s})+r_{\rm s}^2\Phi_{\rm N}^\prime(r_{\rm s})\left(\frac{1}{r}-\frac{1}{r_{\rm s}}\right)\right]\quad r_{\rm s}<r\ll m_c^{-1}.
\end{equation}
In the theories of interest to us we have $\phi_b\ll\phi_c$, which sets the boundary condition $\delta\phi(r_{\rm s})\approx-\phi_c$. Using this in equation \ref{eq:5force}, we find the total force per unit mass (Newtonian plus fifth) in the unscreened region is
\begin{equation}\label{eq:totforce}
 F=F_\phi+F_{\rm N}=\frac{G(r)M(r)}{r^2}\quad \mathrm{where}\quad G(r)=G\left[1+\alpha_c\left(1-\frac{M(r_{\rm s})}{M(r)}\right)\right]
\end{equation}
where $\alpha_c\equiv 2\beta_c^2$ and $M(r)$ is the mass contained inside a shell of radius $r$. Taking the limit as $r\rightarrow\infty$ in equation \ref{eq:phiunscreened} gives us an implicit relation for the screening radius in terms of the Newtonian potential
\begin{equation}\label{eq:screenrad}
 \chi_c\equiv\frac{\phi_c}{2\mpl\beta(\phi_c)}=-\Phi(r_{\rm s})-r_{\rm s}\Phi_{\rm N}^\prime(r_{\rm s}).
\end{equation}
The quantity $\chi_c$ controls the ability of objects to self screen. Clearly if the surface Newtonian potential $\Phi_{\rm N}(R)<\chi_c$ then equation \ref{eq:screenrad} can never be satisfied and the body is completely unscreened ($r_{\rm s}=0$) whilst if the converse is true then the body will be at least partially screened. If the body is completely unscreened then equation \ref{eq:totforce} gives us an order one enhancement of the Newtonian force by a factor $(1+\alpha_c)$.

We can see that whereas $\beta_c$ and $V(\phi_c)$ are the more fundamental of the parameters, it is the combinations $\alpha_c$ and $\chi_c$ which ultimately sets the overall force enhancement and degree of screening and so it is these parameters which our tests of modified gravity constrain. Specific model parameters map into these in straight forward manner and so by working with these combinations we can place constraints in a model independent way. This is extremely useful when the fundamental theories often have more than two parameters, which is often the case. 

In this work we shall not be concerned with demanding that the field also acts as dark energy however, for completeness, we note here that big band nucleosynthesis constraints (BBN) require that the cosmological field remains at the minimum of the effective potential from BBN onwards and so the expansion history is indistinguishable from that of $\Lambda$CDM (structure formation breaks this degeneracy). 

\subsection{Some Common Examples}

\subsubsection{Chameleon Screening}

Any theory where the coupling function $A(\phi)$ is a monotonically increasing function of $\phi$ falls into the class of \textit{chameleon} theories. The chameleon mechanism operates by exploiting the density dependence of the mass of the scalar field; in high density environments the mass is large and the force is very short ranged (Yukawa suppressed) whilst in low density (cosmological) environments the force range can be very large. It is this mass blending in with its environment that gives rise to its name. If the field can indeed reach its minimum at the centre then this large mass ensures that it remains there and varies only in a thin shell near the surface. This has the effect that the force exterior to the object receives contributions from flux lines within this thin shell only and not the entire body and is hence suppressed by a small factor $\Delta R/R\approx\phi_c/6\mpl\beta_c\Phi_{\rm N}(R)$; a phenomena dubbed the \textit{thin shell effect}.

In the original models (Khoury \& Weltman 2004) $\beta(\phi)$ is a constant ($A(\phi)={\rm Exp}[\beta\phi/\mpl]$), however many other models exist in the literature (Gubser \& Khoury (2005); Brax et al. (2010a) and Mota \& Winther (2011)). Our most robust constraints apply to $f(R)$ theories, which are chameleon theories for certain choices of the function $f$ (Brax et al. 2008) with constant $\beta\equiv 1/\sqrt{6}$ so that $\alpha_c\equiv1/3$ and $\chi_c$ arbitrary. Recently, a new unified parametrisation has been discovered (Brax, Davis and Li 2011, Brax et al. 2012) where theories with screening mechanisms can be reverse engineered by specifying a functional form of the cosmological mass and coupling in terms of the scale factor. It would be interesting to repeat our analysis using this unified description and this is left for future work.

There are a wide range of models and parameters that can act as dark energy (Brax et al. 2003; Gannouji et al. 2010; Mota et al. 2011), although some of these have be ruled out using current laboratory searches for fifth forces (see Mota \& Shaw 2006).

\subsubsection{Symmetron Screening}

Symmetron screening (Khoury \& Hinterbichler 2010) relies on symmetry restoration in high density environments to drive the coupling to zero and render the force negligible. 
The simplest model (see Brax et al. 2012 for more general models) uses a $\mathbb{Z}_2$ symmetric potential and coupling function given by
\begin{equation}
 V(\phi)=-\frac{1}{2}\mu^2\phi^2+\frac{\lambda}{4!}\phi^4;\qquad A(\phi)=1+\frac{\phi^2}{2M^2},
\end{equation}
which results in the coefficient of the quadratic term in the effective potential being density dependent:
\begin{equation}
 V_{\rm eff}(\phi)=\frac{1}{2}\left(\frac{\rho}{M^2}-\mu^2\right)\phi^2+\frac{\lambda}{4!}\phi^4.
\end{equation}
In low density environments this coefficient is negative and the field vacuum expectation value (VEV) at the minimum is non-zero-spontaneously breaking the symmetry-whilst in high density environments the symmetry is restored and the VEV moves to zero. The coupling is given at leading order by $\beta\propto\phi$ and so the force enhancement is negligible in high density regions. The requirement that the cosmic acceleration begins in our recent past constrains the parameter $\mu$, resulting in a force range of $\mathcal{O}(Mpc)$ at most and so the symmetron cannot account for dark energy without the inclusion of a cosmological constant (Hinterbichler et al. 2011; Davis et al 2011b).  

\section{TRGB Distances in Modified Gravity}

In this section we briefly show how inferred TRGB distances can be greater than the true value if the core of the star is unscreened. The luminosity of a star ascending the red giant branch is due entirely to a very thin hydrogen burning shell just above the helium core. The triple alpha process ignites in a process known as the helium flash at temperatures $T\sim10^{8}$ K and so for temperatures below this the core has no source of outward pressure and contracts. As the star climbs further the radius of the core decreases whilst the mass increases due to fresh helium being deposited by the shell. This results in a gradual increase until the core temperature is high enough to begin helium burning, at which point the star moves to the left in the H-R diagram. In GR, this sudden jump to the left occurs at a fixed luminosity, however, as we shall see below, this luminosity at the tip is lower in MG, provided that the core is unscreened.

Treating the core as a solid sphere of fixed temperature $T_c$, mass $M_c$ and luminosity $L$, the hydrogen burning shell is incredibly thin and can be treated as having constant mass and luminosity. In this case, the shell pressure and temperature is given by the hydrostatic equilibrium and radiative transfer equations,
\begin{equation}
 \frac{\dd P}{\dd r}=-\frac{GM_c\rho(r)}{r^2};\quad\frac{\dd T^4}{\dd r}=\frac{3}{4a}\frac{\kappa(r)\rho(r)L}{4\pi r^2}
\end{equation}
which can be used to find 
\begin{equation}
 P\propto \frac{G M_cT^4}{L},
 \label{eq:P-T}
\end{equation}
where the opacity in the hydrogen shell is due mainly to electron scattering and so we have taken it to be constant. The pressure in the shell is due mainly to the gas and so we ignore radiation pressure and take the equation of state to be that of an ideal gas, $P\propto \rho T$. Using this and equation \ref{eq:P-T} in the radiative transfer equation we find
\begin{equation}
 T(r)\propto \frac{GM_c}{r},
\end{equation}
where the integration constant is negligible near the base of the shell. Next, we can estimate the luminosity given an energy generation rate per unit mass $\epsilon\propto\rho(r)T(r)^\nu$
\begin{equation}
 L=\int 4\pi r^2\rho(r)\epsilon(r)\dd r.
 \label{eq:lumint}
\end{equation}
For temperatures above $10^7$ K, which is the case in the shell, hydrogen burning proceeds mainly via the CNO cycle and so $\nu=15$. Using the equation of state and the results above in equation \ref{eq:lumint} one finds
\begin{equation}
 L\propto \frac{G^{\frac{8}{3}} M_c^{7.7}} {R_c^6} .
 \label{eq:TRGB}
\end{equation}
 
Now suppose that the core or shell becomes unscreened so that $G(r)\approx G(1+\alpha_e)$ where $\alpha_e$ is the effective value of $\alpha_c$ given by equation \ref{eq:totforce}. The He flash occurs at a fixed temperature, independent of MG, and so if we set $\xi=M_c/R_c$ at the point when $T_c=10^8$ K then we have $\xi_{\rm MG}/\xi_{\rm GR}=(1+\alpha_e)^{-1}<1$. The ratio of the core mass to the core radius at the He flash in MG is then lower than that in GR. In general, this does not tell us anything about the core mass and radius individually, however, in practice one finds that the core radius is the same in both cases (this is borne out by MESA simulations) and so this is a relation between the core masses at fixed temperature. Using equation \ref{eq:TRGB} we then have
\begin{equation}
 \frac{L_{\rm MG}}{L_{\rm GR}}=\frac{1}{(1+\alpha_e)^{5}}
\end{equation}
and hence the peak luminosity is lower in MG, contrary to what one would expect if the argument that radii are generally smaller in MG is followed. 

To infer distances using the TRGB method, one observes the flux ($\propto L/d^2$) and uses the (nearly) universal luminosity at the tip. If the star is indeed unscreened then its luminosity in MG is lower than the universal value assumed, so one would over-estimate the distance.

%
%
\clearpage
\section{Data}

\begin{table}[h]
\begin{center}
\caption{The  galaxies used in the $P-L$ relation and their references. The
second column labelled N gives the number of cepheids observed for each galaxy.
Names that end with * are galaxies with unacceptably large dispersion in P-L relation.}
\bigskip
\begin{tabular}[c]{lll}
Name & N & Reference\\
NGC300 & 117 & Pietrzy{\'n}ski et al. (2002)\\
NGC5253 & 5 & Saha et al (2006)\\
IC4182 & 13 & Saha et al (2006)\\
NGC925 & 79 & Silbermann et al (1996)\\
NGC2541 & 28 & Ferrarese et al. (1998)\\
NGC3319 & 28 & Sakai et al. (1999) \\
NGC1326A & 17 & Prosser et al. (1999)\\
NGC 2090 & 34 & Phelpset al. (1998)\\
NGC 3031 & 25 & Freedman et al. (1994)\\
NGC 3198 & 52 & Kelson et al. (1999)\\
NGC 3351* & 49 & Graham et al. (1997)\\
NGC 3621* & 69 & Rawson et al. (1997)\\
NGC 4321* & 52 & Ferrarese et al. (1996)\\
NGC 4414* & 9 & Turner et al. (1998)\\
NGC 4535 & 50 & Macri et al. (1999)\\
NGC 4548* & 24 & Graham et al. (1999)\\
NGC 4725 & 20 & Gibson et al. (1999)\\
NGC 5457 & 29 & Kelson et al. (1996)\\
NGC 7331 & 13 & Hughes et al. (1998)
\end{tabular}
\end{center}
\label{tab:P-L}
\end{table}

\begin{table}[h]
\begin{center}
\caption{Cepheid and TRGB based distances to the galaxies used in the
paper. The final column gives the screening for $\phi = 4 \times 10^{-7}$ as follows: 
 0: Unscreened, 1: environmentally screened, 2: self-screened.}
\bigskip
\begin{tabular}[c]{llll}
Name & Cepheid D(Mpc) & TRGB D (Mpc) & Screening\\
DDO 069 & 0.71 $\pm$ 0.01 & 0.78 $\pm$ 0.03 & 0 \\
NGC 3109 & 1.15 $\pm$ 0.03 & 1.27 $\pm$ 0.02 & 0 \\
DDO 216 & 1.27 $\pm$ 0.27 & 0.97 $\pm$ 0.03 & 0\\
Sextans A & 1.31 $\pm$ 0.03 & 1.38 $\pm$ 0.03 & 0 \\
Sextans B & 1.49 $\pm$ 0.11 & 1.34 $\pm$ 0.02 & 0 \\
GR8 & 1.80 $\pm$ 0.06 & 2.19 $\pm$ 0.15 & 0 \\
NGC 0300 & 2.03 $\pm$ 0.05 & 1.95 $\pm$ 0.06 & 0 \\
NGC 2403 & 3.20 $\pm$ 0.15 & 3.18 $\pm$ 0.35 & 0\\
NGC 2366 & 3.28 $\pm$ 0.30 & 3.19 $\pm$ 0.41 & 0 \\
NGC 5253 & 3.43 $\pm$ 0.08 & 3.77 $\pm$ 0.19 & 0\\
NGC 4395 & 4.45 $\pm$ 0.37 & 4.61 $\pm$ 0.62 & 0 \\
IC 4182 & 4.68 $\pm$ 0.04 & 4.47 $\pm$ 0.12 & 0 \\
NGC 3621 & 7.17 $\pm$ 0.06 & 7.45 $\pm$ 0.38 & 0\\
SMC & 0.06 $\pm$ 0.00 & 0.06 $\pm$ 0.00 & 1\\
NGC 6822 & 0.51 $\pm$ 0.03 & 0.48 $\pm$ 0.01 & 1\\
IC 1613 & 0.69 $\pm$ 0.01 & 0.72 $\pm$ 0.01 & 1 \\
IC 0010 & 0.72 $\pm$ 0.05 & 0.50 $\pm$ 0.04 & 1\\
M33 & 0.90 $\pm$ 0.02 & 0.88 $\pm$ 0.02 & 1 \\
WLM & 0.95 $\pm$ 0.05 & 0.91 $\pm$ 0.02 & 1 \\
M31 & 0.86 $\pm$ 0.02 & 0.78 $\pm$ 0.02 & 2 \\
NGC 5128 & 3.44 $\pm$ 0.19 & 3.73 $\pm$ 0.24 & 2 \\
M81 & 3.84 $\pm$ 0.06 & 4.04 $\pm$ 0.22 & 2 \\
M83 & 5.01 $\pm$ 0.23 & 4.51 $\pm$ 0.31 & 2 \\
M101 & 7.13 $\pm$ 0.14 & 7.66 $\pm$ 0.39 & 2\\
M106 & 8.41 $\pm$ 0.07 & 7.31 $\pm$ 0.30 & 2 \\
\end{tabular}
\end{center}
\label{tab:distance}
\end{table}

\end{document}